\documentclass[%
 aip,
 amsmath,amssymb,
preprint,%
]{revtex4-1}

\usepackage{dcolumn}
\usepackage{bm}

\usepackage[utf8]{inputenc}
\usepackage[T1]{fontenc}
\usepackage{mathptmx}
\usepackage{etoolbox}

\usepackage{graphicx}
\usepackage{hyperref}
\hypersetup{
    colorlinks=true,
    citecolor=red
}  

\makeatletter
\def\@email#1#2{%
 \endgroup
 \patchcmd{\titleblock@produce}
  {\frontmatter@RRAPformat}
  {\frontmatter@RRAPformat{\produce@RRAP{*#1\href{mailto:#2}{#2}}}\frontmatter@RRAPformat}
  {}{}
}%
\makeatother

\begin{document}


\title{Stochastic fluctuation and transport of tokamak edge plasmas with the resonant magnetic perturbation field}

\author{Minjun J. Choi}
\email{mjchoi@kfe.re.kr}
\author{Jae-Min Kwon} 
\author{Juhyung Kim} 
\author{Tongnyeol Rhee} 
\author{Jun-Gyo Bak} 
\author{Giwook Shin} 
\author{Hyun-Seok Kim} 
\author{Hogun Jhang} 
\author{Kimin Kim} 
\affiliation{Korea Institute of Fusion Energy, Daejeon 34133, Republic of Korea}
\author{Gunsu S. Yun} 
\affiliation{Pohang University of Science and Technology, Pohang, Gyungbuk 37673, Republic of Korea}
\author{Minwoo Kim} 
\affiliation{Korea Institute of Fusion Energy, Daejeon 34133, Republic of Korea}
\author{SangKyeun Kim} 
\affiliation{Princeton University, Princeton, New Jersey 08544, USA}
\author{Helen H. Kaang} 
\affiliation{Korea Institute of Fusion Energy, Daejeon 34133, Republic of Korea}
\author{Jong-Kyu Park} 
\affiliation{Princeton Plasma Physics Laboratory, Princeton, New Jersey 08543, USA}
\author{Hyung Ho Lee} 
\author{Yongkyoon In} 
\author{Jaehyun Lee} 
\author{Minho Kim} 
\author{Byoung-Ho Park} 
\affiliation{Korea Institute of Fusion Energy, Daejeon 34133, Republic of Korea}
\author{Hyeon K. Park} 
\affiliation{Ulsan National Institute of Science and Technology, Ulsan 44919, Republic of Korea}

\begin{abstract}

We present that a statistical method known as the Complexity-Entropy analysis is useful to characterize a state of plasma turbulence and flux in the resonant magnetic perturbation (RMP) edge localized mode (ELM) control experiment. 
The RMP ELM suppression phase with the stochastic pedestal top temperature fluctuation can be distinguished from the natural ELM free phase with the chaotic fluctuation. 
It is discussed that the stochastic temperature fluctuation localized near the pedestal top can be originated from the narrow layer of the field penetration near the pedestal top. 
The forced magnetic island can emit the resonant drift wave of comparable sizes (relatively low-k) in the RMP ELM suppression phase, and it can results in the generation of stochastic higher wavenumber fluctuations coupled to tangled fields around the island.  
The analysis of the ion saturation current measurement around the major outer striking point on the divertor shows that it also becomes more stochastic as the stronger plasma response to the RMP field is expected.

\end{abstract}

\maketitle

\section{Introduction}
\label{sec:int}

The external magnetic perturbation field has been utilized for various purposes in tokamak plasma experiments including the control of serious plasma magnetohydrodynamic instabilities.
For example, the resonant magnetic perturbation (RMP) field is the most recognized method for the suppression or mitigation of an edge localized mode (ELM) instability~\cite{Evans:2004fm, Liang:2007fx, Jeon:2012hua, Sun:2016ez, Suttrop:2016ci, Park:2018ex} which is driven by the large current density and the steep pressure gradient of the edge pedestal region in the high confinement mode (H-mode) tokamak plasmas.
Quasi-periodic collapse of the pedestal due to the explosive growth of the ELM is a critical issue for steady-state operation of the H-mode plasmas.
With the penetration of the RMP field, the pedestal pressure can be maintained at the lower level than the the ELM stability threshold, i.e. the edge transport is enhanced to avoid the ELM growth and the pedestal collapse.

Although understanding of the magnetic perturbation field effects on plasma transport has been significantly enhanced recently during the pursuit of the most efficient and reliable control of the ELM instability, it is a long-standing complicated nonlinear problem and further researches are required at the fundamental level including plasma turbulence characteristics. 
One prominent feature of the RMP ELM suppressed pedestal is the increase of plasma turbulence observed in frequency or wavenumber spectral measurements~\cite{McKee:2013je, Lee:2016ku, Sung:2017ie}.
However, spectral analyses provide the limited information such as the local wavenumber or the phase velocity in the laboratory frame, and it is often not trivial to characterize a pedestal state based on spectral measurements of plasma turbulence. 

In this work, we used a statistical method known as the Complexity-Entropy analysis~\cite{Rosso:2007vb} to distinguish a state of plasma turbulence and to extract more information. 
This statistical analysis was shown to be useful to tell a deterministic chaotic signal (generated by the logistic map, for example) from a stochastic signal (of fractional Brownian motion, for example): throughout this paper we restrict the meaning of \textit{stochastic} or \textit{chaotic} as given by the Complexity-Entropy analysis results (see below).
Broadband electron temperature fluctuations, or plasma turbulence, are measured near the H-mode pedestal top of KSTAR plasmas, and how its statistical characteristics changes with the RMP field is investigated using the Complexity-Entropy analysis.  

Since the ultimate reason for the RMP ELM control is to avoid the divertor erosion and the statistical characteristics of divertor particle flux is of great interest~\cite{Morales:2011gl, Garcia:2012jo}, the ion saturation current measurement obtained using the Langmuir probe around the major striking point is also analyzed across different levels of the RMP field penetration. 

The remainder of this paper is organized as follows.
In section \ref{sec:exp}, a typical RMP ELM suppression plasma in KSTAR and the key diagnostics are introduced briefly, and the Complexity-Entropy analysis we adapted is explained. 
Results and discussion on the analysis of the pedestal temperature turbulence and the divertor particle flux are provided in section \ref{sec:res}. 
We found that the Complexity-Entropy analysis is able to characterize a state of plasma turbulence and transport in the RMP ELM control experiment.
Conclusion is provided in section \ref{sec:con}.
The characterization of turbulence and transport would be the foundation for understanding of plasma confinement in general configurations other than the RMP ELM control experiment in tokamaks.    

\section{Experimental set-up and method}
\label{sec:exp}

\subsection{Edge localized mode suppression by the resonant magnetic perturbation field in KSTAR}

\begin{figure}[t]
\centering
\includegraphics[keepaspectratio,width=0.49\textwidth]{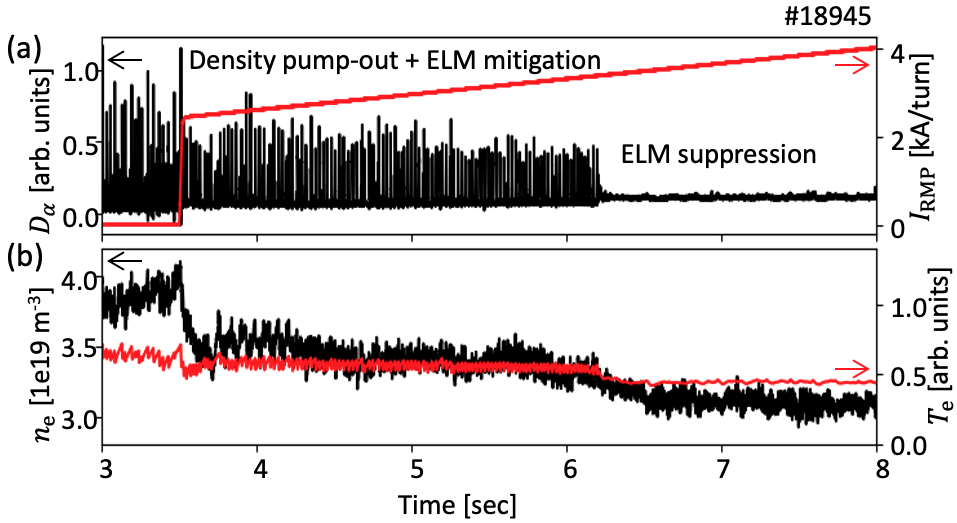}
\caption{(Color online) (a) The $D_{\mathrm{\alpha}}$ emission (black) and the coil current for the resonant magnetic perturbation (RMP) field $I_\mathrm{RMP}$ (red). The edge localized mode (ELM) is completely suppressed above some RMP field strength threshold. (b) The line averaged plasma density $n_{\mathrm{e}}$ (black) and the pedestal electron temperature $T_{\mathrm{e}}$ (red).}
\label{fig:ree}
\end{figure}

An example of the RMP ELM suppression observed in the KSTAR H-mode deuterium plasma \#18945 is shown in Fig. \ref{fig:ree}. 
The $n=1$ RMP field is applied from $t=3.5$~sec where $n$ represents the toroidal mode number.
Its strength increases in time as indicated by the coil current (red line) shown in Fig. \ref{fig:ree}(a). 
The ELM pedestal collapse is mitigated with the RMP field and it is completely suppressed after $t \sim 6.2$~sec as shown by the quiescent divertor $D_{\mathrm{\alpha}}$ emission signal whose sharp rise indicates the pedestal collapse event. 
Fig. \ref{fig:ree}(b) shows that the line averaged plasma density $n_{\mathrm{e}}$ drops about 17~\% with applying the RMP field while the edge electron temperature $T_{\mathrm{e}}$ drops about 9~\%, i.e. the so-called density pump-out. 
Absolute $n_{\mathrm{e}}$ and $T_{\mathrm{e}}$ measurements are obtained by two color interferometry diagnostics~\cite{Lee:2016fw} and the electron cyclotron emission diagnostics~\cite{Jeong:2010bq}, respectively. 
Further decrease of $n_{\mathrm{e}}$ and the edge $T_{\mathrm{e}}$ during an abrupt transition from the ELM mitigation to the ELM suppression is observed, indicating that the pedestal pressure height is more reduced to make the ELM stable. 

\subsection{Fluctuation and particle flux diagnostics}

\begin{figure}[t]
\centering
\includegraphics[keepaspectratio,width=0.49\textwidth]{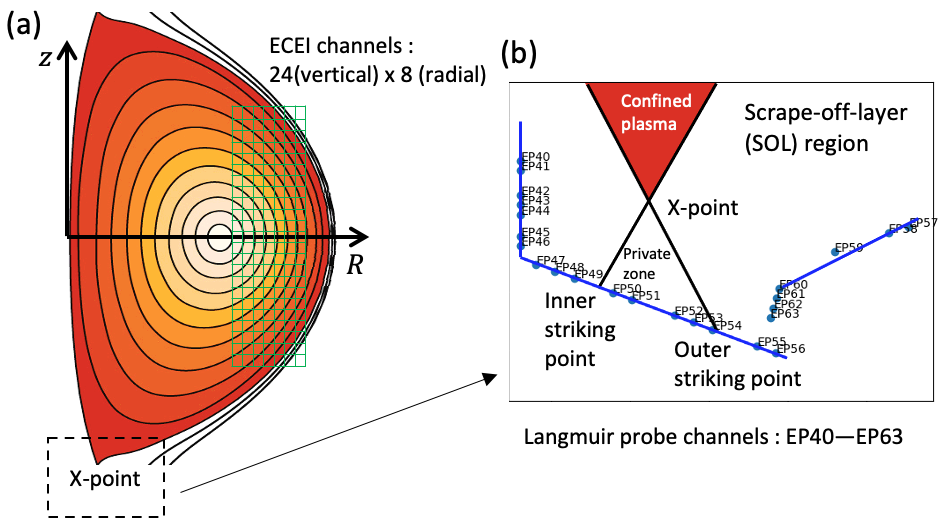}
\caption{(Color online) (a) Channel distribution of the electron cyclotron emission imaging (ECEI) diagnostic in two-dimensional real space ($R$,$z$). (b) Channel distribution of the divertor Langmuir probes around the lower X-point of the KSTAR plasma.}
\label{fig:el}
\end{figure}

The local electron temperature measurement near the pedestal top is obtained using the electron cyclotron emission imaging (ECEI) diagnostics on the KSTAR tokamak~\cite{Yun:2014kv, Park:2019kq}.
It consists of 24 (vertical) $\times$ 8 (radial) channels, and their distribution in two-dimensional ($R$, $z$) space is illustrated in Fig. \ref{fig:el}(a). 
It can measure the local $T_\mathrm{e}$ with a high spatial ($\Delta R \approx \Delta z =$ 1--2 cm) and temporal ($\Delta t =$ 0.5--2 $\mu$s) resolution.

The particle flux can be estimated by the ion saturation current measurement using the Langmuir probe. 
High spatial resolution measurements across the outer striking point was obtained by the slow back and forth movements of plasma against one of Langmuir probes (see Fig. \ref{fig:el}(b)) for a quasi-stationary state.    

\subsection{The Complexity-Entropy analysis}

This analysis is intended for evenly sampled serial data such as a time series with the fixed sampling rate, and it is shown to tell a chaotic signal generated from a deterministic system from a stochastic signal~\cite{Rosso:2007vb}.
The distinction relies on the probability distribution of amplitude orders in partial segments of the signal~\cite{Pompe:2002jv}. 
For example, the first segment of size $d=3$ of a serial data $x = \{x_i\}_{i=1,...,N} = \{4,7,9,6,\cdots,x_N\}$ is $\{4,7,9\}$, and the order of its elements is simply increasing and can be represented by $(1,2,3)$. 
The second (overlapping) segment is $\{7,9,6\}$ and the order is $(3,1,2)$.
The probability distribution of these amplitude orders is called the Bandt-Pompe (BP) probability distribution~\cite{Pompe:2002jv}.
Using the BP probability of a given data $P=\{p_j\}_{j=1,...,d!}$ where subscript $i$ represents each amplitude order (there are $d!$ possible ways of the amplitude ordering for the segment size $d$. Note that $N$ should be large enough than $d!$ for the reliable calculation of the BP probability~\cite{Pompe:2002jv}.), the Jensen Shannon Complexity $C_{\mathrm{JS}}$ and the normalized Shannon Entropy $H$, which are the basis of the Complexity-Entropy analysis~\cite{Rosso:2007vb}, can be calculated as follows.
The Jensen Shannon Complexity ($C_{\mathrm{JS}} = Q H $) is defined as the product of the normalized Shannon Entropy $H=S/S_{\mathrm{max}}$ and the Jensen Shannon divergence $Q$. 
$S=S(P)=-\sum_{j} p_j \ln(p_j)$ is the Shannon Entropy of the given BP probability distribution $P=\{p_j\}$ and $S_{\mathrm{max}}$ is the maximum possible Entropy, i.e. $S_{\mathrm{max}} = \ln(d!)$ with the equiprobable distribution $P_{\mathrm{e}} = \{p_j\} = 1/d!$.
The Jensen Shannon divergence is given as $Q = Q_0 \left\{ S\left(\frac{P+P_{\mathrm{e}}}{2}\right) - \frac{S(P)}{2} - \frac{S(P_{\mathrm{e}})}{2} \right\}$ where $Q_0$ is the normalization constant $Q_0 = -2 / (\frac{d! + 1}{d!} \ln(d! + 1) - 2\ln(2 d!) + \ln(d!))$~\cite{Martin:2006dz, Maggs:2015ff}. 

Both Entropy and Complexity, in this paper, are information theoretic (Shannon) terminologies and have specific meanings.
Entropy means a measure of missing (unknown) Information of the given probability distribution.
It has the maximum for the equiprobable distribution from which one can learn almost nothing about the system.
Complexity was suggested as the product of Disequilibrium and Entropy (missing Information) to capture the intuitive notion about a complex system~\cite{RuizPLA1995}, where Disequilibrium is a measure of distance from the equiprobable distribution.
The Jensen Shannon divergence is used for Disequilibrium for the Complexity-Entropy analysis~\cite{Rosso:2007vb}.
The idea of Complexity can be understood by two examples~\cite{RuizPLA1995} of simple (not complex) system in physics, i.e. a perfect crystal and an ideal gas.
A perfect crystal, represented by a peaked probability distribution of states, has very small missing Information but large Disequilibrium, and their product Complexity will remain small. 
On the other hand, an ideal gas, represented by the equiprobable distribution of states, has large missing Information but very small Disequilibrium, and their product Complexity also will remain small.

In reference~\cite{Rosso:2007vb}, it was shown that signals from various chaotic systems (the logistic map, the skew tent map, Henon's map, etc) and signals from stochastic processes (fractional Brownian motion (fBm) or fractional Gaussian noise (fGn)) can be separated in the Complexity-Entropy plane. 
Following researches show that this method is effective to characterize density fluctuations in fusion plasma experiments~\cite{Maggs:2013fv, Maggs:2015ff, Zhu:2017fo}.

Motivated by previous studies, we adapted the Complexity-Entropy analysis as follows.
The boundary between chaotic and stochastic signals in the Complexity-Entropy plane is the locus of points of fBm and fGn (see Figure 4 of reference~\cite{Maggs:2013fv}). 
Signals far above this locus is called chaotic, and signals close to or below this locus is called stochastic. 
For the efficient comparison of a given data against fBm or fGn, the Jensen Shannon Complexity $C_{\mathrm{JS}}$ is rescaled as 
\begin{equation}
\hat{C}=\frac{C_{\mathrm{JS}} - C_0}{|C_{\mathrm{bdry}} - C_0|}
\end{equation}
where $C_0(H)$ is the Jensen Shannon Complexity of fBm or fGn and $C_{\mathrm{bdry}}(H)$ is the maximum (if $C_{\mathrm{JS}} > C_0$) or minimum (if $C_{\mathrm{JS}} < C_0$)~\cite{Calbet:2001eq, Zhu:2017fo} Jensen Shannon Complexity at the given $H$. 
The rescaled complexity ($\hat{C}$) ranges from -1 ($C_{\mathrm{JS}} = C_{\mathrm{min}}$) to 1 ($C_{\mathrm{JS}} = C_{\mathrm{max}}$), and the less $\hat{C}$ means the less chaotic or the more stochastic in the relative comparison.

\section{Results and discussion}
\label{sec:res}

\subsection{Analysis of temperature fluctuation near the pedestal top with resonant magnetic perturbation field}

\begin{figure}[t]
\centering
\includegraphics[keepaspectratio,width=0.49\textwidth]{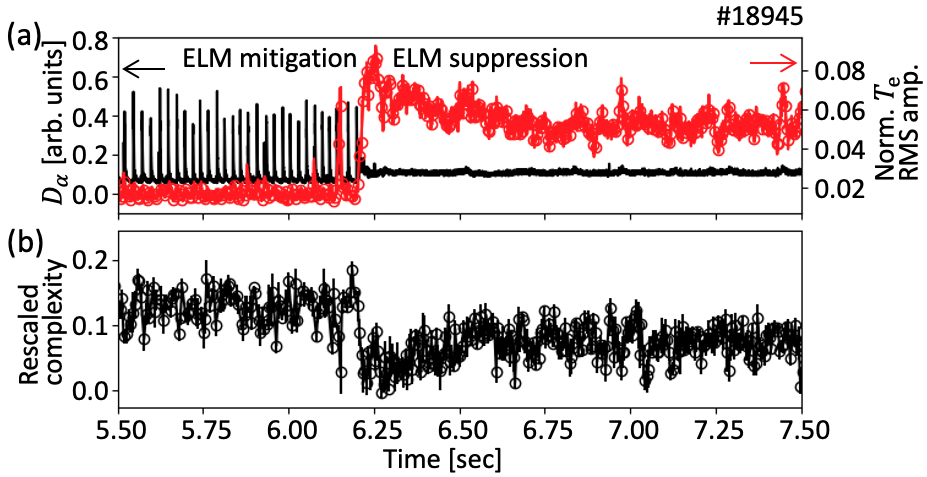}
\caption{(Color online) (a) The $D_{\mathrm{\alpha}}$ emission (black) and the root mean square (RMS) amplitude of the normalized electron temperature $T_\mathrm{e}$ fluctuation near the pedestal top (red) (b) The rescaled complexity of the temperature fluctuation near the pedestal top. Vertical error bars indicate the standard deviation of measurements.}
\label{fig:ir}
\end{figure}

Increase of broadband density and temperature fluctuations has been observed in the edge region of the RMP ELM suppressed H-mode plasmas~\cite{McKee:2013je, Lee:2016ku, Sung:2017ie}.
In KSTAR experiment, the increased fluctuation is often localized near the pedestal top as shown in Fig.~\ref{fig:ir}(a). 
The $T_{\mathrm{e}}$ fluctuation amplitude at the pedestal top ($R_\mathrm{ped}$) increases with the abrupt transition to the RMP ELM suppression from the RMP ELM mitigation, and it remains at a higher level during the suppression phase.
Local measurement of the $T_{\mathrm{e}}$ fluctuation is provided by the ECEI diagnostics~\cite{Yun:2014kv} around the pedestal top~\cite{Yun:2012dd} which is identified by the strongest ELM precursor location. 
A separated study found that this strongest ELM precursor location coincides with the pedestal top in the ion temperature profile within the measurement uncertainty. 
The root mean square (RMS) amplitude of the normalized $T_{\mathrm{e}}$ fluctuation is obtained by integrating 0--100 kHz of the short time (5 ms) cross power spectra between vertically adjacent ECEI channels to suppress the noise contribution~\cite{Choi:2019tw}. 

The rescaled complexity of the pedestal top $T_{\mathrm{e}}$ fluctuation is calculated to characterize a turbulence structure recorded in $T_{\mathrm{e}}$ temporal variation. 
Parameters for the BP probability calculation are set as $d=5$ ($3 \le d \le 7$ is typically suggested) and $N=2500$ which satisfy $N/d! > 20$.
Since the time of interest for turbulence analysis is the transit time of a turbulence structure across the ECEI channel (a few centimeter / a few kilometers per second), i.e. a few microseconds, it requires the fast sampling rate of the diagnostics data. 
The ECEI diagnostics in our experiments has the sufficiently fast sampling rate (the time step $\Delta t = 2$~us), and sequential data segments of $d \Delta t = 10$~us length in the 5~ms window ($N=2500$) were used to get one BP probability distribution.
The length of segments, on the other hand, restricts that the Complexity-Entropy analysis is intrinsically more sensitive to the fluctuation whose frequency is close to 100 kHz ($=1/(d \Delta t)$).
It is found empirically that the fluctuation whose frequency is less than 20 kHz does not affect the analysis result much. 
Also, note that the moments of pedestal collapses by ELMs are excluded from the analysis.    

The analysis result shows that the rescaled complexity $\hat{C}$ decreases with the ELM suppression transition and remains at a lower level during the suppression phase as shown in Fig.~\ref{fig:ir}(b). 
The reduced $\hat{C}$ means that the measured temperature fluctuation signal near the pedestal top becomes more stochastic in the ELM suppression phase.

\subsubsection{Comparison with the natural edge localized mode free case}

\begin{figure}[t]
\centering
\includegraphics[keepaspectratio,width=0.49\textwidth]{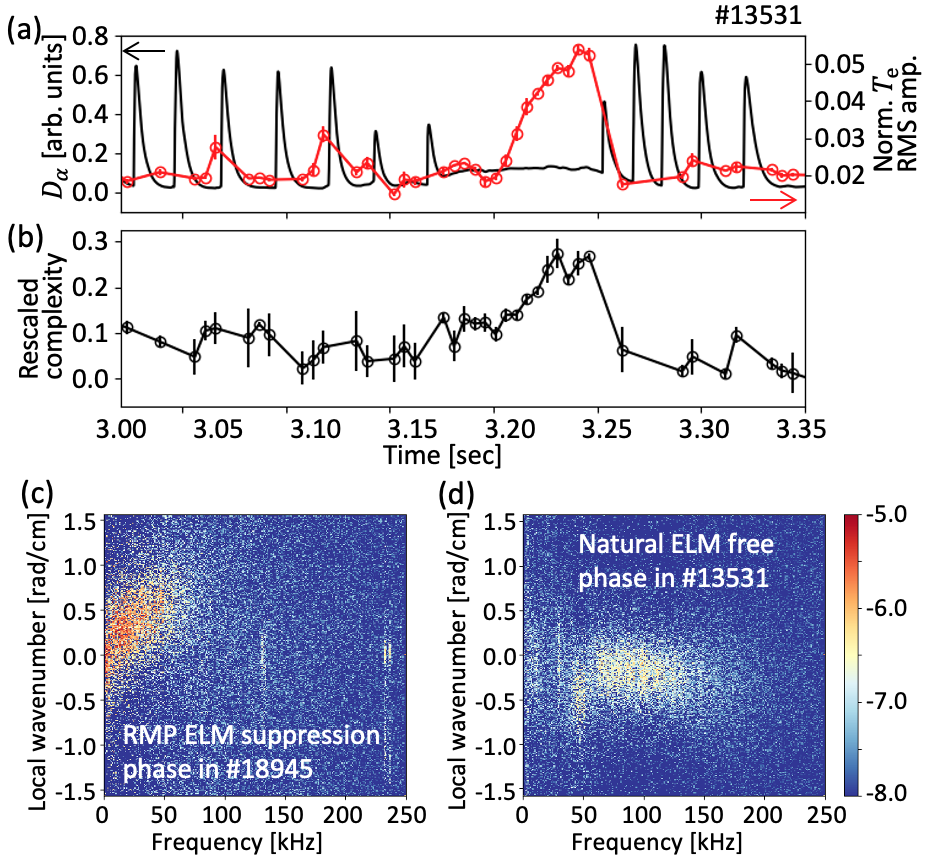}
\caption{(Color online) (a) The $D_{\mathrm{\alpha}}$ emission (black) and the root mean square (RMS) amplitude of the normalized electron temperature $T_\mathrm{e}$ fluctuation near the pedestal top (red) (b) The rescaled complexity of the temperature fluctuation near the pedestal top. Vertical error bars indicate the standard deviation of measurements. The local wavenumber frequency spectra of the temperature fluctuation in (c) the RMP ELM suppression phase and (d) the natural ELM free phase.}
\label{fig:cwn}
\end{figure}

One may think that the reduced rescaled complexity $\hat{C}$ is simply a result of the increased broadband fluctuation rather than the RMP field effects.
For comparison, the broadband $T_{\mathrm{e}}$ fluctuation signal near the pedestal top from a natural ELM free phase without the external perturbation field (\#13531) was analyzed in the same way as the previous RMP ELM suppression analysis. 
The short natural ELM free phase appears for 3.20--3.25~sec with the significant increase of the broadband (0--170 kHz) $T_{\mathrm{e}}$ fluctuation amplitude as shown in Fig.~\ref{fig:cwn}(a).   
The local wavenumber frequency spectra using the poloidally adjacent ECEI channels~\cite{Beall:1998fx, Choi:2019tw} found that the increased $T_{\mathrm{e}}$ fluctuations in both the \#18945 RMP ELM suppression phase (Fig.~\ref{fig:cwn}(c)) and the \#13531 natural ELM free phase (Fig.~\ref{fig:cwn}(d)) have shown the similar broadband feature ($k_\theta \rho_\mathrm{i} < 1$ where $k_\theta$ is the poloidal wavenumber and $\rho_\mathrm{i}$ is the ion Larmor radius). 
However, the $\hat{C}$ of the $T_{\mathrm{e}}$ fluctuation near the pedestal top increases in correlation with the the $T_{\mathrm{e}}$ fluctuation amplitude during the natural ELM free phase as shown in Fig.~\ref{fig:cwn}(b), which is in contrast with the RMP ELM suppression phase result. 
This shows that the Complexity-Entropy analysis can be used to characterize the measured signal of turbulence and the reduction of the rescaled complexity in the RMP ELM suppression phase does not simply result from the increase of broadband fluctuation (see below for comparison between RMP ELM suppression phases). 
Also, Fig.~\ref{fig:cwn}(b) itself shows that turbulence develops to have a more chaotic structure in the natural ELM free phase without the perturbation field. 

\subsubsection{Hypothesis and supporting observation}

The natural following question is how the magnetic perturbation field affects the statistical characteristics of turbulence. 
Plasma response to the perturbation field can be either kink or tearing, and both might affect indirectly (via changes in pressure/flow profile) or directly (via mode coupling) turbulence characteristics.
Our hypothesis is that the RMP field penetration near the pedestal top and the resulting tangled magnetic field changed turbulence structure more stochastic through direct scattering or coupling. 
This is inspired by two recent studies: (1) The numerical simulation using TM1~\cite{Yu:2011pr} suggested that the narrow penetration of the perturbation field at the pedestal top is responsible for the ELM suppression~\cite{Hu:2020ik}. (2) The theoretical study showed that turbulence structure can lock on to the structure of tangled magnetic fields~\cite{Cao:2021}. 

It is not trivial to demonstrate the localized penetration of the perturbation field (or the resulting tangled magnetic fields) around the H-mode pedestal top since the local measurement of magnetic fields is not possible in that region.  
In the literature, the penetration of the magnetic fields was acknowledged by the electron temperature profile flattening~\cite{Nazikian:2015bg} or the reduction of the turbulence correlation length~\cite{Xu:2006dg}. 

\begin{figure}[t]
\centering
\includegraphics[keepaspectratio,width=0.49\textwidth]{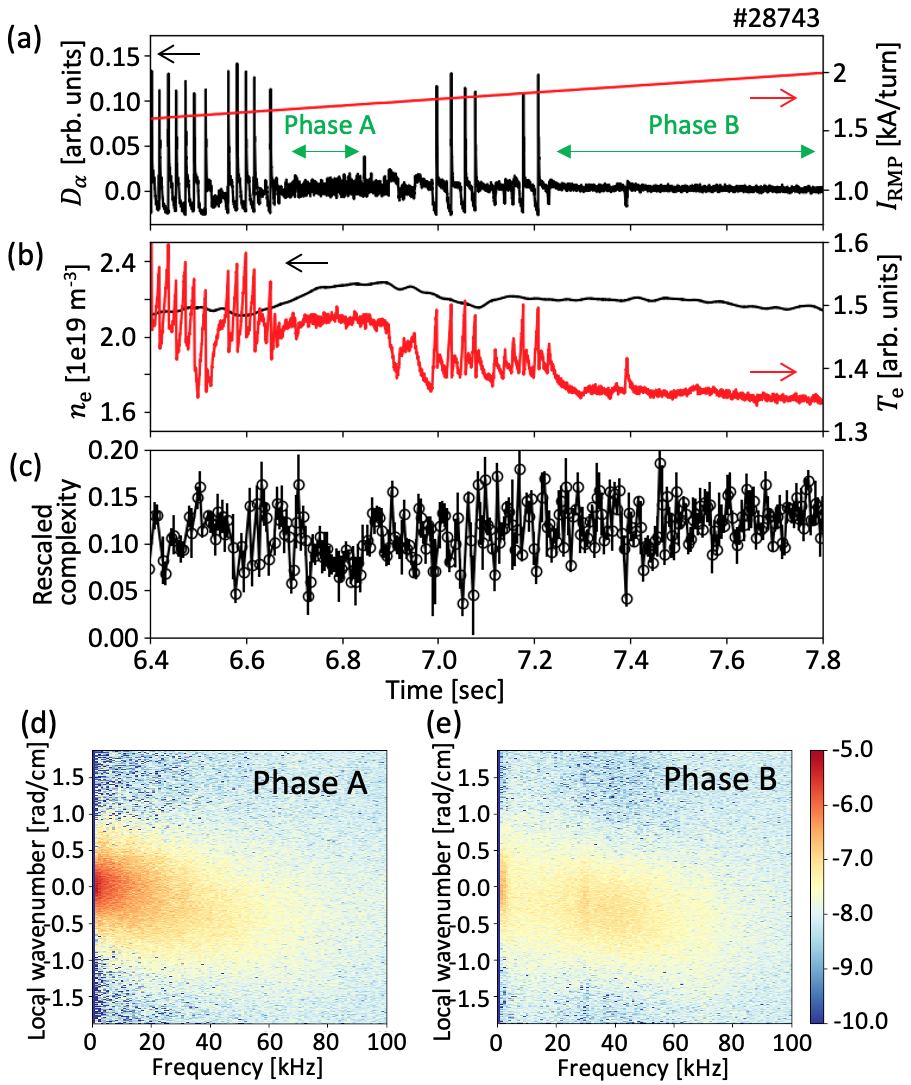}
\caption{(Color online) (a) The $D_{\mathrm{\alpha}}$ emission (black) and the coil current for the resonant magnetic perturbation (RMP) field $I_\mathrm{RMP}$ (red). Two phases A and B of the RMP edge localized mode (ELM) suppression are indicated. (b) The line averaged plasma density $n_{\mathrm{e}}$ (black) and the pedestal electron temperature $T_{\mathrm{e}}$ (red). (c) The rescaled complexity of the temperature fluctuation near the pedestal top. Vertical error bars indicate the standard deviation of measurements. The local wavenumber frequency spectra of the temperature fluctuation in (d) the phase A and (e) the phase B.}
\label{fig:cab}
\end{figure}

\begin{figure}[t]
\centering
\includegraphics[keepaspectratio,width=0.4\textwidth]{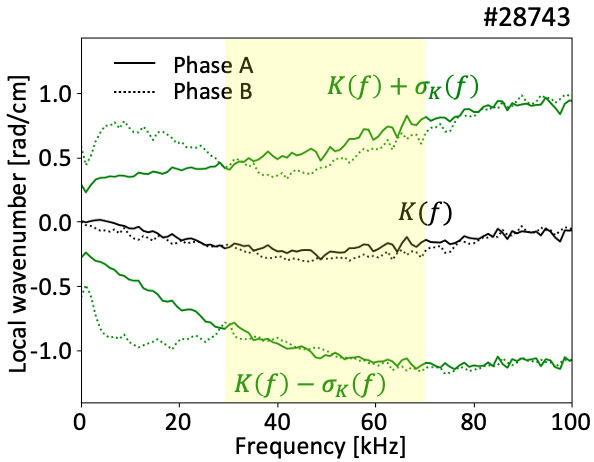}
\caption{(Color online) The estimated local wavenumber $K(f) = \sum_k k s(k,f)$ and spectral width $\sigma_K(f) = \sqrt{\sum_k (k- K(f))^2 s(k,f)}$ using the local wavenumber frequency spectra $s(k,f)$ in the phase A (bold) and the phase B (dotted).}
\label{fig:cl}
\end{figure}

Although the measurement of the sufficiently fine electron temperature profile was not available in KSTAR experiments, a discharge where the reduction of the correlation length is correlated with the rescaled complexity $\hat{C}$ reduction was found. 
The increasing $n=1$ RMP field is applied in the KSTAR experiment \#28743, and two distinguished RMP ELM suppression phases (A and B) are observed as shown in Fig.~\ref{fig:cab}.  
The $\hat{C}$ of the pedestal top $T_{\mathrm{e}}$ fluctuation is calculated in time, and it is lower in the phase A (around 6.8 sec) than the phase B (after 7.3 sec) as shown in Fig.~\ref{fig:cab}(c). 
The local wavenumber frequency spectra of the pedestal top $T_{\mathrm{e}}$ fluctuation in Figs.~\ref{fig:cab}(d) and \ref{fig:cab}(e) show that the effective fluctuations for $\hat{C}$ calculation ($f > 20$~kHz) are comparable between two phases. 
The correlation length $l_\theta$ of the $T_{\mathrm{e}}$ fluctuation can be estimated by the inverse of the wavenumber spectral width $\sigma_K$ of the local wavenumber frequency spectra $s(k,f)$~\cite{Beall:1998fx, Xu:2006dg}. 
Fig.~\ref{fig:cl} shows the local wavenumber $K(f) = \sum_k k s(k,f)$ (black) and $K(f) \pm \sigma_K(f)$ (green) where $\sigma_K(f) = \sqrt{\sum_k (k- K(f))^2 s(k,f)}$ in the phase A (bold) and the phase B (dotted). 
Since the $\sigma_K(f)$ estimation~\cite{Beall:1998fx} strongly depends on the signal-to-noise ratio, it can be properly compared between two phases in the frequency regime (30--70 kHz, highlighted in Fig.~\ref{fig:cl}) where the signal-to-noise ratio is comparable.
In that regime, the correlation length, the inverse of $\sigma_K(f)$, is shorter in the phase A than the phase B.
The shorter correlation length of the pedestal top $T_{\mathrm{e}}$ fluctuation in the phase A suggests that tangled magnetic fields by the field penetration are more stronger near the pedestal top in the phase A~\cite{Xu:2006dg}.
The lower $\hat{C}$ of the $T_{\mathrm{e}}$ fluctuation in the phase A may be consistent with our hypothesis that the $\hat{C}$ can decrease by tangled magnetic fields by the field penetration.    

\subsubsection{Comparison with the previous TM1 simulation prediction}

\begin{figure}[t]
\centering
\includegraphics[keepaspectratio,width=0.49\textwidth]{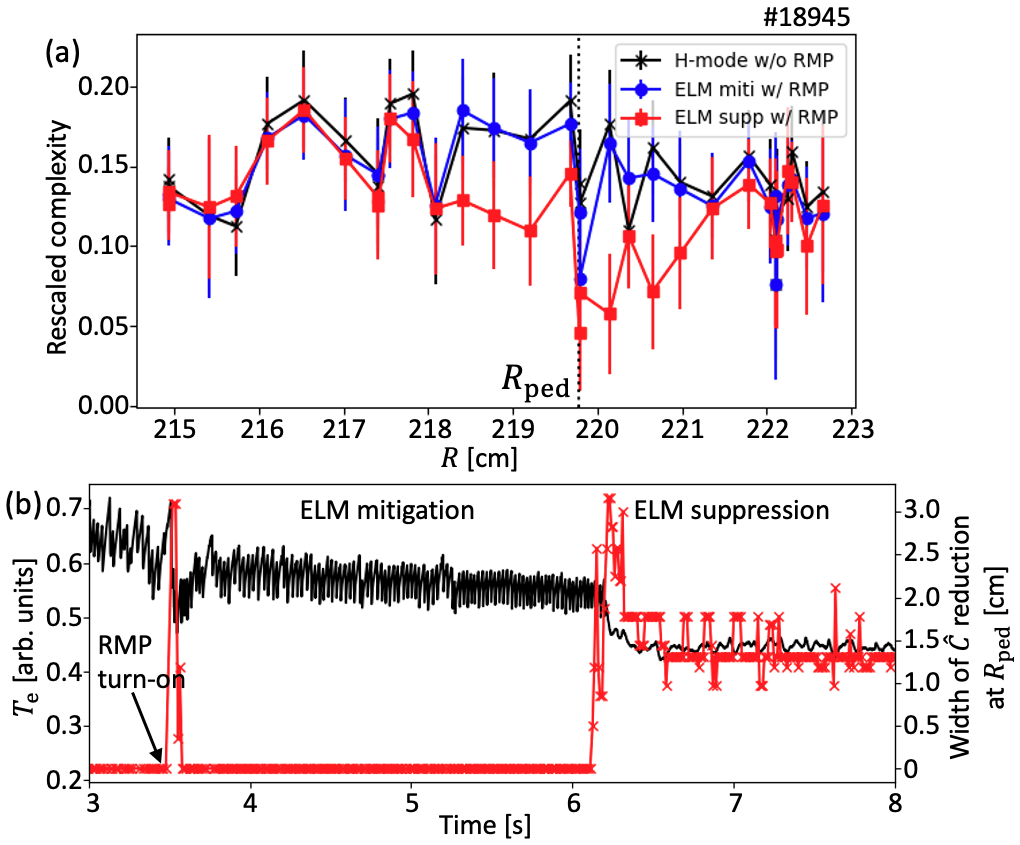}
\caption{(Color online) (a) Radial profiles of the rescaled complexity $\hat{C}$ of electron temperature $T_\mathrm{e}$ fluctuation in different phases. The pedestal top location $R_\mathrm{ped}$ is indicated by the vertical dotted line. Error bars indicate the standard deviation of measurements in each phase, and measurements for the moments of pedestal collapses by edge localized modes are excluded. (b) The pedestal $T_{\mathrm{e}}$ (black) and the full width of the $\hat{C}$ reduction region (red crosses).}
\label{fig:rw}
\end{figure}

Postulating that the rescaled complexity $\hat{C}$ of the $T_{\mathrm{e}}$ fluctuation can reflect effects of the field penetration, its reduction near the pedestal top with the transition to the ELM suppression from the mitigation in Fig.~\ref{fig:ir}(b) (or in the phase A of Fig.~\ref{fig:cab}(c)) means that the ELM is suppressed with the field penetration near the pedestal top. 
This interpretation is line with the result of the aforementioned TM1 simulation~\cite{Hu:2020ik}.  
Specifically, the simulation suggested that the sequential formation of the narrow field penetration layers, which are not overlapping, separately at the pedestal foot and top could explain the density pump-out in the ELM mitigation phase and the ELM suppression, respectively. 
To identify the $\hat{C}$ variation layers and assess its narrowness, the radial $\hat{C}$ profiles of the $T_{\mathrm{e}}$ fluctuation in different phases are plotted in Fig.~\ref{fig:rw}(a). 
The $\hat{C}$ values are obtained from the two-dimensional ($R, z$) $T_{\mathrm{e}}$ fluctuation measurements, and they are projected on the midplane ($z=0$) radial axis including the plasma center based on the reconstructed equilibrium using EFIT~\cite{Lao:1985hn}. 
Since each channel has different noise contribution, the relative change of profiles between different phases is more meaningful than their absolute value. 
The $\hat{C}$ profile is nearly same between the H-mode phase without the RMP field (black crosses) and the ELM mitigation phase with the RMP field (blue circles). 
This is inconsistent with the simulation result which expects the formation of the field penetration layer around the pedestal foot with the ELM mitigation to explain the density pump-out.
However, the ECE measurement around the pedestal foot region is not reliable due to the limited diagnostics capability with the low density and temperature of that region, and so it should be carefully taken. 
On the other hand, as expected, a significant drop compared to values of other periods is observed near the pedestal top $R_\mathrm{ped}$ in the initial ELM suppression period with the RMP field (red squares). 

The full width of the $\hat{C}$ reduction region can be estimated as the range over which the $\hat{C}$ decreases more than the standard deviation (error bars) of $\hat{C}$ measurements in each period.
The evolution of the $\hat{C}$ reduction width is shown in Fig.~\ref{fig:rw}(b) for the long period.  
When the RMP field is first applied around 3.5~sec and the electron temperature drops momentarily, the $\hat{C}$ reduction over a significant region is observed for that moment. 
Note that the correlation between the $\hat{C}$ reduction and the electron heat transport event can be another indirect supporting observation for the close link between the field penetration and the $\hat{C}$ reduction.   
The width decreases back to the zero level when the plasma enters to the ELM mitigation phase with the partially recovered pedestal probably with the field screening. 
It increases to a significant level with the transition to the ELM suppression phase. 
The full width is about $2.6 \pm 2.0$~cm in the initial ELM suppression phase and about 1.2~cm in the later stationary phase.
Considering that the typical width of the H-mode pedestal in KSTAR plasmas is about 2~cm and that the flux surface distortion by the plasma kink response can cause the overestimation of the $\hat{C}$ reduction width, it may be sufficiently small, i.e. the half width of the $\hat{C}$ reduction < the pedestal width, to call it a localized layer especially for the stationary suppression phase. 

\subsubsection{Discussion on the RMP ELM suppression}

According to previous analyses, the localized field penetration near the pedestal top may be important for entering the ELM suppression phase, though it would not be the only way: the phase B in Fig.~\ref{fig:cab}, for which the field penetration near the pedestal top may not be significant, seems to represent a different path for the ELM suppression. 
Interestingly, the field penetration condition expected from the two fluid MHD theory, i.e. $\omega_{\perp,e} = \omega_{*e} + \omega_{E} \sim 0$, seems not to be satisfied for the ELM suppression phase in DIII-D experiments~\cite{Hu:2020ik}. 
Here, $\omega_{\perp,e}$ is the electron perpendicular flow frequency, $\omega_{*e}$ is the electron diamagnetic drift frequency, and $\omega_E$ is the electric drift frequency.  
This might imply that there is a missing step in between the field penetration and the ELM suppression. 

The ELM suppression phase from experiments is often characterized by (1) the increased temperature or density fluctuation and (2) the $\omega_E \sim 0$ condition. 
In addition, our analyses imply that (3) the field penetration layer (or, the size of the forced magnetic island) can be narrow.  
These remind us of the idea of the drift wave emission by the nonlinear resonance between a small magnetic island and drift waves~\cite{Waelbroeck:2001dl}. 

\begin{figure}[t]
\centering
\includegraphics[keepaspectratio,width=0.49\textwidth]{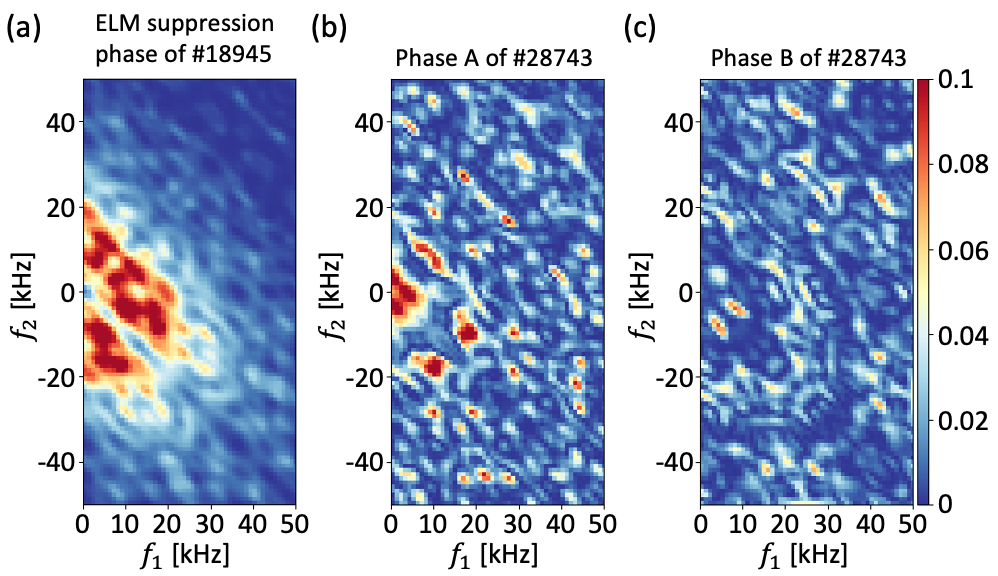}
\caption{(Color online) The auto squared bicoherence of the pedestal top electron temperature fluctuation in different phases. (a) The edge localized mode (ELM) suppression phase by the resonant magnetic perturbation (RMP) field in \#18945. The RMP ELM suppression (b) phase A and (c) phase B in \#28743.}
\label{fig:bic}
\end{figure}

Before checking the condition for the nonlinear resonance~\cite{Waelbroeck:2001dl}, the (auto squared) bicoherence is calculated using the pedestal top $T_{\mathrm{e}}$ fluctuation to first confirm the existence of the phase coupled components in the fluctuation. 
The bicoherence is a well known method to detect the strength of the phase coupling among triples of fluctuations of frequencies ($f_1$, $f_2$, $f_3=f_1+f_2$)~\cite{Kim:1979ps}. 
Fig.~\ref{fig:bic} shows that the bicoherence at $f_3 = f_1+ f_2$ in different phases. 
The RMP ELM suppression phases with the rescaled complexity reduction, which may be indicative of tangles fields around a magnetic island, have significant bicoherence mostly in low frequencies ($\le 20$~kHz) as shown in Figs.~\ref{fig:bic}(a) and \ref{fig:bic}(b), while the significant fluctuation triple coupling is not observed in another ELM suppression phase without the $\hat{C}$ reduction as shown in Fig.~\ref{fig:bic}(c). 

Then, it can be discussed if the phase coupled low frequency (low-k) fluctuations could result from the nonlinear resonance with the presumable forced magnetic island. 
The key nonlinear resonance condition for the drift wave emission can be summarized as follow~\cite{Waelbroeck:2001dl}.
\begin{eqnarray}
\omega_E < \omega_\mathrm{island} < \omega_{*e}
\end{eqnarray}
where $\omega_\mathrm{island}$ is the island frequency in the laboratory frame. 
For a forced magnetic island by the static RMP field, $\omega_\mathrm{island} \sim 0$, and considering DIII-D measurements in the ELM suppression phase~\cite{Hu:2020ik}, i.e. $\omega_E \sim 0$ and $\omega_{*e} \neq 0$, it would be possible for the nonlinear resonance condition to be satisfied. 
The island would emit drift waves of the radial wavelength $\lambda_{\mathrm{DW}}$ similar to its radial width $W_\mathrm{island}$~\cite{Waelbroeck:2001dl}. 

\begin{figure}[t]
\centering
\includegraphics[keepaspectratio,width=0.49\textwidth]{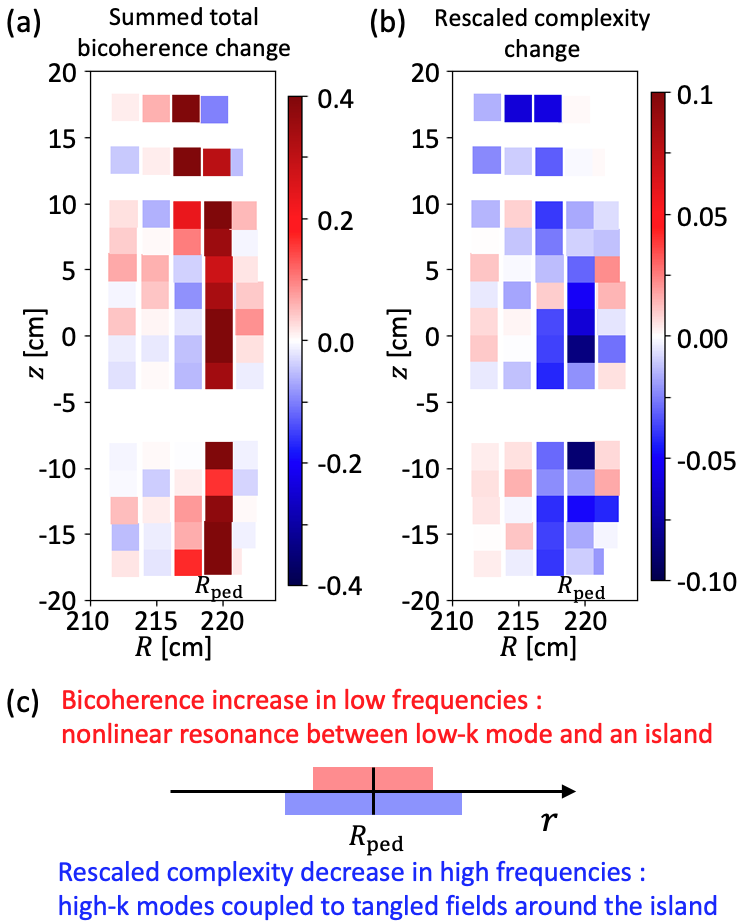}
\caption{(Color online) Changes of (a) the summed total bicoherence and (b) the rescaled complexity of electron temperature fluctuation between the ELM mitigation and initial suppression phases in \#18945. The pedestal top location $R_\mathrm{ped}$ is around $R=220$~cm. (c) An illustration to understand the bicoherence increase in low frequency fluctuations and the rescaled complexity decrease originated from high frequency fluctuations by a small magnetic island near the pedestal top and its tangled fields, respectively.}
\label{fig:br}
\end{figure}

To estimate the size of the presumable magnetic island by the bicoherence increase, the bicoherence change between the ELM mitigation and the initial ELM suppression phases of \#18945 is analyzed using two-dimensional $T_{\mathrm{e}}$ measurements in the edge region. 
Fig.~\ref{fig:br}(a) shows that the summed total bicoherence increases locally near the pedestal top. 
Comparing with two-dimensional measurements of the rescaled complexity change shown in Fig.~\ref{fig:br}(b), the bicoherence increment region seems to be narrower than the rescaled complexity reduction region (see the midplane measurements).
Then, the island size might be about a few $\rho_i$, and the drift waves are in the measurable radial wavelength range of the KSTAR ECEI diagnostics and might correspond to the measured phase coupled fluctuation.  
Note that some channels whose noise contributions are exceptional or signals are saturated are excluded in this analysis.

Based on above analyses and discussion, we arrive at one picture (Fig.~\ref{fig:br}(c)) to understand our observations.
Firstly, a forced small magnetic island forms near the pedestal top by the field penetration~\cite{Hu:2020ik}.
Next, the resonance condition for the drift wave emission~\cite{Waelbroeck:2001dl} is satisfied to generate relatively low-k (low frequency) fluctuations: explaining the bicoherence increase. 
These localized low frequency fluctuations might enhance the heat flux to make the pedestal height slightly lower than the ELM threshold. 
On the other hand, high-k (high frequency) fluctuations can be coupled to tangled magnetic fields around the island~\cite{Cao:2021}: explaining the rescaled complexity decrease (more stochastic fluctuation).    
This view stresses the importance of the coupled evolution between a magnetic island and turbulent fluctuations~\cite{Ishizawa:2019ky, Ida:2019jn, Choi:2021fs, Choi:2021rm} in the RMP ELM suppression. 

\subsection{Analysis of the diverter particle flux with resonant magnetic perturbation field} 

The statistical characteristics of the divertor particle or heat flux would be important to understand the edge-SOL transport as well as to develop the flux model~\cite{Morales:2011gl, Garcia:2012jo} for the divertor erosion study. 
In KSTAR, measurements of the ion saturation current ($\sim$ the particle flux) around the major striking point are obtained by the divertor Langmuir probe when the striking point drifts across the probe location (Fig.~\ref{fig:el}(b)).
The data from three plasmas with different RMP levels, i.e. \#19023 (H-mode without the RMP field), \#19130 (ELM mitigation with the RMP field), and \#18945 (ELM suppression with the RMP field) are analyzed to study the RMP field effect. 
The BP probability calculation parameters are set to $d=5$ and $\Delta t = 1$~us to capture the short pulses in the Langmuir probe data~\cite{Zhu:2017fo}.
Note that all the data could be obtained using the same Langmuir probe in the near term so that the absolute value of the rescaled complexity $\hat{C}$ can be directly compared assuming the similar noise contribution.
In fact, the base level of rescaled complexity measurements in the private zone has shown a similar value in different plasmas. 

\begin{figure}[t]
\centering
\includegraphics[keepaspectratio,width=0.7\textwidth]{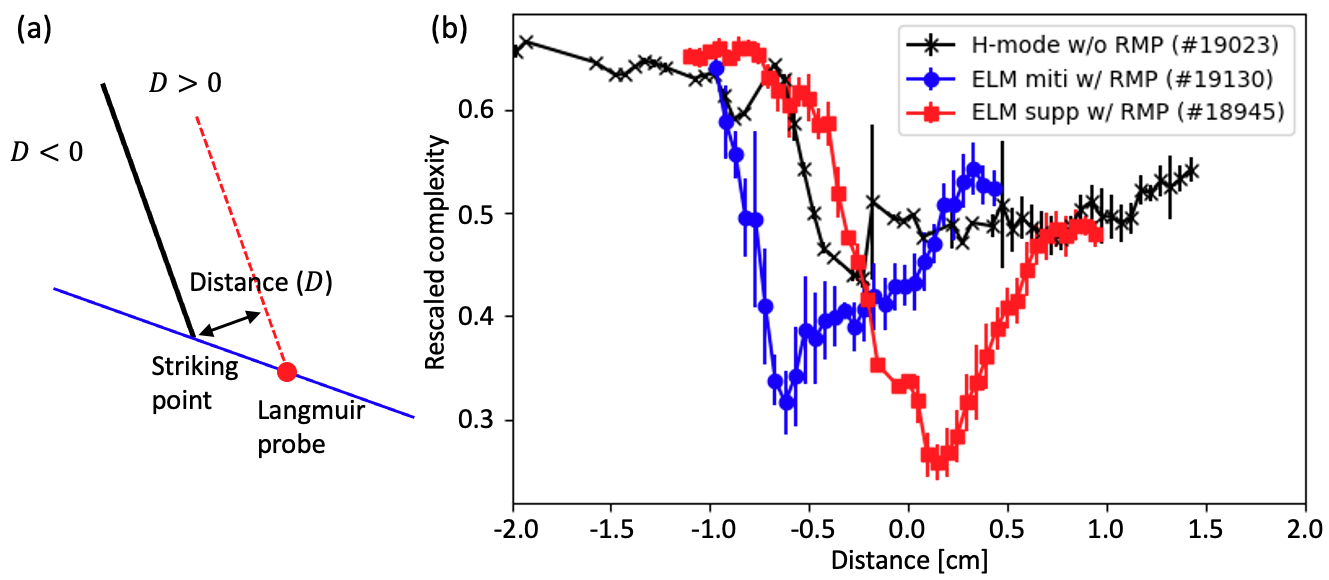}
\caption{(Color online) (a) The illustration for the distance $D$ measurement between flux surfaces of the outer striking point (black dot) and the Langmuir probe (red dot). $D$ can have a different offset for each plasma due to the equilibrium reconstruction uncertainty with the perturbation field. The ion saturation current measurements along $D$ are obtained as the striking point moves across the Langmuir probe with the plasma movement. (b) Radial profiles of the rescaled complexity of the ion saturation current around the striking point in different phases. The same Langmuir probe is used for absolute comparison of the profiles. Error bars indicate the standard deviation of measurements.}
\label{fig:lrp}
\end{figure}

The $\hat{C}$ profiles of the Langmuir probe data along the distance $D$, which is defined as a distance between two flux surfaces containing the major outer striking point and the probe (see Fig.~\ref{fig:lrp}(a)), are shown in Fig.~\ref{fig:lrp}(b). 
A field line tracing code~\cite{Rhee:2019ns} is used to obtain the flux surface at the probe location. 
By the definition, $D=0$ would correspond to the striking point, $D<0$ to the private zone, and $D>0$ to the typical scrape-off-layer region, respectively, but $D$ estimation can involve about 1~cm absolute error due to the equilibrium reconstruction uncertainty with the RMP field.
The $\hat{C}$ profiles from different plasmas have a similar shape, i.e. it drops fast from $D \ll 0$ towards $D=0$ and increases slowly over the $D>0$ region. 
The fast drop might correspond to a transition from the private zone to the striking point. 
Interestingly, the minimum of $\hat{C}$, which is expected to reside close to the major outer striking point, is found to be significantly different with different RMP levels. 
The lower minimum is achieved, i.e. the ion saturation current (particle flux) becomes more stochastic, as the stronger plasma response is expected. 
This sequential $\hat{C}$ reduction might be associated with the sequential RMP field penetration observed in the numerical simulation~\cite{Hu:2020ik}.
Caution that, however, three plasmas had different pedestal parameters such as $q_{95}$ and the different RMP field configurations, and further analysis based on data from the controlled experiments will be required.

\section{Conclusion} 
\label{sec:con}

Using the Complexity-Entropy analysis, a state of plasma turbulence and flux under the RMP field can be characterized. 
The analyses suggest that the rescaled complexity of the local temperature fluctuation can reflect the degree of tangled magnetic fields by the field penetration. 
In addition, based on the analyses and previous numerical~\cite{Hu:2020ik} and theoretical~\cite{Waelbroeck:2001dl, Cao:2021} researches, the turbulence state during one RMP ELM suppression phase can be understood as summarized in Fig.~\ref{fig:br}(c). 
The forced magnetic island~\cite{Hu:2020ik} near the pedestal top can emit the resonant drift wave of comparable sizes (relatively low-k)~\cite{Waelbroeck:2001dl} in the RMP ELM suppression phase, and it can results in the generation of stochastic high-k fluctuations coupled to tangled fields around the island~\cite{Cao:2021}.
The analysis of the ion saturation current measurement shows that a stochastic model of the flux should be used in the situation where the strong plasma response to the RMP field is expected.   
  
\section*{Acknowledgments}

The authors would like to express sincere gratitude to the KSTAR team. 
M.J.C. acknowledges helpful discussion with M. Cao and P.H. Diamond during the 6th Asia-Pacific Transport Working Group meeting. 
This research was supported by R\&D Programs of ``KSTAR Experimental Collaboration and Fusion Plasma Research(EN2201-13)" and ``High Performance Fusion Simulation R\&D(EN2241-8)" through Korea Institute of Fusion Energy (KFE) funded by the Government funds and by National Research Foundation of Korea under NRF-2019M1A7A1A03088462.

\section*{Competing interests}

The authors declare no competing interests.

\section*{Data availability}

Raw data were generated at the KSTAR facility. 
Derived data are available from the corresponding author upon request. 

\section*{Code availability}

The data analysis codes used for figures of this article are available via the GitHub repository~\url{https://github.com/minjunJchoi/fluctana}~\cite{Choi:2019tw}. 

\bibliographystyle{naturemag}


\end{document}